# A novel model-based equation for size dependent mean recovery coefficients for spheres and other shapes

Robin de Nijs[1,2]

1. Department of Clinical Physiology and Nuclear Medicine, Copenhagen University Hospital, Rigshospitalet, Copenhagen, Denmark
2. Neurobiology Research Unit, Copenhagen University Hospital, Rigshospitalet, Copenhagen, Denmark

**Running title**: A novel equation for size dependent mean recovery coefficients

**Keywords**: recovery coefficient, SPECT, PET, phantom, simulation, spheres, ellipsoids, surface area-to-volume ratio, dosimetry, partial volume effect

<u>Corresponding author</u>
Robin de Nijs, orcid.org/0000-0002-3908-3013
Department of Clinical Physiology and Nuclear Medicine
Copenhagen University Hospital, Rigshospitalet
Blegdamsvej 9
DK-2100, Copenhagen, Denmark
Phone: +45-35454011
E-mail: robin.de.nijs@regionh.dk

**Abstract**

*Background*: In NM-imaging, theoretical curves for the recovery coefficient (*RC*) of the signal maximum and mean are known for spheres and cubes, if a 3D Gaussian PSF is assumed. The *RC* of the maximum is also known for cylinders. For these and other shapes empirical equations with one or two fit-parameters have been utilized.

*Methods*: An equation for the *RC* for large objects of arbitrary shape is derived and generalized into an empirical equation for smaller objects, which is verified by numerical simulations. The proposed equation is compared to published results on SPECT kidney phantom measurements and to PET measurements on the NEMA IEC PET body phantom with six spheres.

*Results*: The signal loss (1-*RC*) for large spheres is inversely proportional to the radius, where the slope is proportional to the FWHM of the spatial resolution. For non-spherical shapes the generalized instead of the volume equivalent radius should be utilized. For smaller objects, an equation with one added empirical fit-parameter is presented. It is demonstrated that the EANM-guidelines' two-parameter logistic function results in a poor fit if the theoretical slope and inverse proportionality are forced and it gives a suboptimal fit when both parameters are fitted.

*Conclusions*: A novel model-based equation for the mean *RC*-curve is derived. It can be used for arbitrary shapes as long as the sphericity is taken into account and it is accurate down to *RC*=10%. One parameter is directly related to the spatial resolution, while the other is a shape depending fit-parameter.

## Introduction

Due to limited spatial resolution of imaging systems in nuclear medicine (NM) such as SPECT (Single Photon Emission Computed Tomography) and PET (Positron Emission Tomography) images are blurred resulting in spatial broadening of especially small objects, constituting the so-called partial volume effect (PVE). The total measured signal inside an object originates from both the radioactivity inside and outside of the object. The signal originating from within the object will be partly located outside the object due the limited spatial resolution causing a spill-out of signal. In a similar way radioactivity outside the object gives rise to a signal inside the object, which is called spill-in. The PVE consists of the combination of both spill-in and spill-out and depends on both object size and shape. The PVE is not caused by limited voxel size, even though insufficient voxel size can also result in signal loss, when the object is comparatively small.

### *Definitions and effect of the background activity*

One way of correcting for the PVE is the use of recovery coefficients (*RC*). The *RC* is defined as the ratio between measured and known activity concentration [1][2][3]. The contrast recovery coefficient (*CRC*) is a similar ratio, where the uniform background is subtracted from the concentrations[1][4][5]. These coefficients are defined by

$$RC = \frac{C'}{C} \qquad \text{and} \qquad CRC = \frac{C' - C'_{\mathrm{BG}}}{C - C_{\mathrm{BG}}}, \tag{1}$$

where $C$ is the activity concentration within the object and $C_{\mathrm{BG}}$ the activity concentration in the background. Measured concentrations are denoted by an apostrophe. The recovery coefficient is sometimes defined as a background corrected *RC*[5][6] as *CRC* in equation (1). The two definitions in equation (1) are related by

$$RC = (1 - B)CRC + \frac{C'_{\mathrm{BG}}}{C_{\mathrm{BG}}} B \qquad \text{with} \quad B = \frac{C_{\mathrm{BG}}}{C}, \tag{2}$$

where $B$ is the known background to object activity concentration. For a hot background ($B$>1) the $CRC$-definition still yields correct values between zero and one, but the $RC$-definition does not, and the cold sphere recovery coefficient[4] should be employed instead.

If the system's calibration factor is determined by a large volume of interest in the background, as often is the case, the measured and known background concentrations have the same value, which is also theoretically the case, when only the influence of the limited spatial resolution is considered. Equal values for measured and known background concentrations are assumed in Kessler et al[4] and in the following.

Kessler et al[4] showed theoretically that $CRC$ is independent of the background concentration, but Sakaguchi et al[7] showed that for larger values ($\gtrsim$0.2) of the background ratio $B$ the full width half maximum (FWHM), which expresses the observed spatial resolution, increased significantly. Note that $RC$ cannot be lower than the value for $B$ according to equation (2) and that the $RC$ equals one for $B$=1. This does not agree with the findings of Raskin et al[8], but is in agreement with Staanum[9]. However, Stenvall et al[10] showed that the relationship between $B$ and $RC$ might not be linear.

The recovery coefficients in equation (1) can be utilized for different volumes as is the case for the Standard Uptake Value (SUV)[3][11]. The latter is defined as a local activity concentration scaled by the average activity concentration in the entire body, and therefore the numerator should be corrected for the PVE by utilizing a recovery coefficient. $CRC_{\mathrm{max}}$ and $RC_{\mathrm{max}}$ are defined by equation(1) for the maximum value of the concentrations, while $CRC_{\mathrm{mean}}$, and $RC_{\mathrm{mean}}$ also are defined by equation (1), where the mean value of the concentrations in the object is applied. Measurements of the maximum SUV and $(C)RC_{\mathrm{max}}$ are often noise sensitive due to the measurement of the maximum and may depend on the voxel size. The $RC_{\mathrm{mean}}$ is also equal to the fraction of the total amount of activity that is recovered. In the following, we will assume there is zero background and denote the zero background $CRC_{\mathrm{mean}}$=$RC_{\mathrm{mean}}$ as the recovery coefficient $RC$.

*Empirical and theoretical equations for RC-curves*

A common method of experimentally investigating the recovery coefficient is measuring a phantom with spheres of different sizes. Often the NEMA IEC PET body phantom with six spheres is used. The spheres have an inner volume within the range from 0.5 to 26.5 ml, and an inner diameter of 10, 13, 17, 22, 28 and 37 mm. Volumes of Interest (VOI) are typically drawn based on CT (Computed Tomography) or knowledge of the phantom design, resulting in a recovery coefficient as a function of size. Unfortunately, the *RC* is sometimes plotted as a function of volume - and not of radius - resulting in a curve featuring a very sharp decreasing slope and a plateau for large volumes[12]. Recently, Grings et al[13] experimentally showed that for printed three-dimensional (3D) kidney models, the surface area-to-volume ratio *S:V* is closely proportional to the signal loss (1-*RC*). For a sphere this ratio is inverse proportional to its radius. Therefore, *RC*s should be plotted as a function of radius *r* (or better yet its reciprocal) and not volume *V*.

According to Ramonaheng et al[2] the *RC*-curve is fitted with

$$RC = b_1 - b_2 \mathrm{e}^{-b_3 r}. \tag{3}$$

with three fit-parameters $b_1$, $b_2$ and $b_3$. It is important to note that the *RC* should equal unity for infinitely large objects. Therefore, the first fit parameter in equation (3) is set to unity ($b_1$=1), which reduces the number of fit parameters to two.

According to the EANM-guidelines on dose calculations[14] the *RC*-curve is fitted by a volume dependent logistic equation

$$RC = 1 - \frac{1}{1+\left(\frac{V}{\alpha}\right)^{\beta}} = \frac{1}{1+\left(\frac{\alpha}{V}\right)^{\beta}} \tag{4}$$

with fit-parameters $\alpha$ and $\beta$. The mathematically equivalent expression on the right hand side can be found in Tran-Gia et al[12]. Alternatively, this equation can be written as a function of diameter [15][16][17] or radius instead of volume. If proportionality (of the signal loss) to *S:V* is forced[13] for large objects ($V \to \infty$ or $r \to \infty$) EANM's equation (4) should have $\beta$=⅓. This demand cannot be met by equation (3), but it is fulfilled by a mono-exponential equation with fit-parameter *k*:

$$RC = e^{-\frac{k}{r}} . \qquad (5)$$

An analytical approach was published by Kessler et al[4]. It assumes a 3D Gaussian Point Spread Function (PSF). An analytical equation for both spheres and cylinders was derived for the position dependent signal. In order to calculate the mean *RC*, Kessler's rather complicated equations need to be integrated over the object volume. This has recently been achieved for spheres by Gabiña et al[18]. The *RC* for the position with the highest signal, denoted as $RC_{max}$, can be found by using Kessler's equations at the origin or by integration of the PSF, where the limits of integration are the object boundaries. Kessler's equations for $RC_{max}$ are given in Appendix I supplemented with the $RC_{max}$ equation for a cube. Di Martino et al[19] derived *RC*-equations for cubic shapes. Unfortunately, equation (8) in their paper leading to $RC_{mean}$ is not entirely correct. The correct equation for $RC_{mean}$ for a cube is given in Appendix I as well.

In the case of a small object $RC_{max}$ can be used as an estimate for $RC_{mean}$, since only the maximum value is contained in the object. Equations (3-5) will be compared to the simulated *RC*-curves. It is important to distinguish between *RC* and $RC_{max}$, because (for large objects) the signal loss is proportional to the reciprocal size, while $RC_{max}$ remains constant and is close to one. Therefore, the known statement that the partial volume effect is only relevant in case of sizes of more than three times the system's FWHM[20] only holds true for $RC_{max}$.

## Methods

### *Recovery coefficients*

For large objects, *RC* can be estimated by recognizing that the signal loss 1-*RC* is proportional to the surface area *S* and a smoothing thickness $\delta$, which depends on the spatial resolution $\sigma$. For large 3D-objects the signal loss can be approximated by

$$1 - RC = \frac{S \cdot \delta}{V}, \qquad (6)$$

where *V* is the volume of the object. The product of the surface area and the thickness represents a volume of original intensity containing the lost signal. In order to calculate the smoothing thickness, the case of a 1D semi-infinite source is considered, which is represented by a Heaviside step function $H$, representing the transition at $x$-position $x$=0 between the object at $x$>0 with intensity one and the background at $x$<0 with intensity zero. The step function is convolved with a 1D Gaussian PSF yielding an expression for the 1D image intensity $I(x)$:

$$I(x) = \int_{-\infty}^{\infty} H(x') \cdot PSF(x - x')dx' = \frac{1}{\sigma\sqrt{2\pi}} \int_{0}^{\infty} e^{-\frac{(x-x')^2}{2\sigma^2}} dx' \quad (7)$$

The signal loss (spill-out) is given by the area under $I(x)$ for $x$<0. This area also equals the product of the smoothing thickness times the step height, which is one. The smoothing thickness $\delta$ is therefore given by

$$\delta = \int_{-\infty}^{0} I(x)dx = \frac{1}{\sigma\sqrt{2\pi}} \int_{-\infty}^{0} \int_{0}^{\infty} e^{-\frac{(x-x')^2}{2\sigma^2}} dx'\, dx = \frac{\sigma}{\sqrt{2\pi}} \quad (8)$$

The evaluation of the integral can be found in Appendix II. In the case of a semi-infinite 3D slab as an approximation for large objects the smoothing thickness from equation (8) can be used in equation (6).

In order to take the influence of different shapes into account, the generalized radius $\rho$[21], the volume equivalent radius $r$[22], and the sphericity $\psi$[23] are utilized and these quantities are related by

$$\rho = \frac{3V}{S} = \psi r \qquad \text{with} \qquad \psi = \frac{S_{\text{sphere}}}{S} = \frac{\sqrt[3]{36\pi V^2}}{S}, \quad (9)$$

where $S_{\text{sphere}}$ is the surface area of a sphere with the same volume as the object. The volume equivalent radius $r$ is defined by the radius $r$ of a sphere with the same volume as the object. The sphericity $\psi$ is one for spheres and less than one for other shapes. Assuming that the smoothing thickness derived for the 1D case can be applied to a large 3D object, combining equations (6,8,9) results in

$$1 - RC \approx \frac{\sigma}{\sqrt{2\pi}} \cdot \frac{S}{V} = \frac{3}{\sqrt{2\pi}} \cdot \frac{\sigma}{\rho} = \frac{c}{r} \qquad \text{with} \qquad c = \frac{3}{\sqrt{2\pi}} \cdot \frac{\sigma}{\psi}. \quad (10)$$

This equation is valid for large objects of any shape, for which the generalized radius is significantly larger than the system resolution $\sigma$. Equation (10) is equivalent to equation (9) in Gabiña et al[18], which was derived for large spheres, and it is also in agreement with the linear relationship between $RC$ and surface area-to-volume ratio as observed by Grings et al[13]. Rewriting of equation (10) as a function of FWHM yields

$$1 - RC = \frac{\text{FWHM}}{4\sqrt{\pi}\ln 2} \cdot \frac{S}{V} = 1.0165 \cdot \frac{\text{FWHM}}{2\rho} = \frac{0.8193}{\psi} \cdot \frac{\text{FHWM}}{V^{1/3}}$$

$$\text{with} \quad \text{FWHM} = \sigma \cdot \sqrt{8\ln 2}. \tag{11}$$

In Table 1 some objects are listed with their generalized radius and sphericity. For different organs, typical values for the sphericity can be determined as is done for kidney phantoms[13]. Intersubject variation of the sphericity is expected to be small or negligible.

Equation (10) gives a numerical value for the slope of the *RC* curve for large objects, which can be used in the equations (4) and (5) mentioned earlier. This means that the only fitting parameter left is the spatial resolution $\sigma$ of the system.

For smaller objects, the *RC*-curve is no longer described by equation (10). According to the latter equation *RC* is zero for $\sigma/\rho \approx 0.84$, but in reality *RC* will be larger than zero. An empirical equation that is equal to equation (10) for large objects and that describes the curve bending behavior for smaller objects is given by

$$RC = \frac{a}{e^{\frac{ac}{r}} + a - 1} \approx 1 - \frac{c}{r}, \tag{12}$$

where *c,* which is directly related to the systems' spatial resolution $\sigma$, is defined in equation (10) and *a* is an empirical fit-parameter. The last approximation is valid for large objects, and it is identical to equation (10) as is required. For *a*=1 equation (12) reduces to the mono-exponential in equation (5) with *k*=*c*.

*Numerical simulations*

For spheres, cubes, cylinders and ellipsoids, see Table 1, numerical simulations were performed in Matlab R2021a (MathWorks Inc., Natick, Massachusetts, USA). Shapes were defined by a 3D mask with voxel values of one inside the object and zero outside the object. This mask *M* was convolved by a 3D Gaussian kernel, resulting in a blurred image *I.* Convolution was performed by multiplication of the 3D Fast Fourier Transform (FFT) of mask and kernel in frequency space, and taking the inverse 3D FFT of the product. Matrix size *n* was 256 pixels in each direction. Sampling considerations regarding maximum and minimum FWHM can be found in Appendix III. Fitting of the curves was performed with the Levenberg-Marquardt nonlinear least squares algorithm. The recovery coefficient was determined as the ratio between the voxelwise summations of the masked image and the mask itself:

$$RC = \frac{\sum M \cdot I}{\sum M} \,. \tag{13}$$

Previous simulations were published by Soret et al[20], but unfortunately the diameter was not scaled with the FWHM, resulting in different curves for identical FWHM to diameter ratios. Also, the *RC* was not plotted against the reciprocal value of the diameter, which would have revealed the linear behavior for large objects.

*Phantom measurements*

In order to investigate the practical application of equation (10) published kidney phantom data acquired with $^{99m}$Tc SPECT[13] was fitted and analyzed. Background has an influence on measured recovery coefficients, so therefore the subset with zero background, called "Infinity Target to Background Ratio", was selected. The subset for SPECT-reconstructions with 72 iterations was used, since Grings et al[13] only reported regression analysis for this subset. Based on a linear fit and measured sphericity values, *RC*-values were calculated and compared to the measured *RC*-values. These calculated *RC*-values are referred to as "Grings". The average sphericity for the cortex-only and whole-parenchyma phantoms was calculated and used to calculate fit-based *RC*-values as well. The latter *RC*-values are referred to as "fixed", because of the fixed values for the sphericity.

Also, data acquired with the NEMA IEC PET body phantom with six spheres for accreditation according to both the $^{18}$F standard 1 and 2 in EANM's EARL-program for $^{18}$F PET[3] was analyzed. The EARL-datasets were acquired on a Siemens Biograph Vision 600 PET/CT (Siemens, Knoxville, TN, USA). Due to the EARL-definition of the recovery coefficient as *RC* in equation (1) the background needs to be taken into account. The theoretical *CRC* is used to calculate the theoretical *RC* by using equation (2) with $B \approx 0.1$. Measured and known background concentration are assumed to be identical. EARL-data was reconstructed with four iterations and five subsets ("4i5s") without (EARL1) and with (EARL2) point spread function modeling.

## Results

*Numerical simulations*

First, the proposed proportionality with the reciprocal size according to equation (10) was investigated. In Figure 1 the recovery coefficients for the shapes in Table 1 are plotted as a function of the generalized radius and the equivalent spherical radius. The radius on the horizontal axis in Figures 1 is written as $R$, which can refer both to generalized and volume equivalent radius. Up to a ratio between $\sigma$ and radius of approximately 0.25 ($RC \approx 0.7$), equation (10) is in good agreement with the simulated values, when the generalized radius is used. The $RC$-curve based on the volume equivalent spherical radius deviates for all non-spherical shapes. This is in agreement with the phantom experiments and theoretical curve for a sphere, which can be found in the paper of Gabiña et al[18].

In Figure 2, the $RC$ for spheres is shown as a function of the reciprocal generalized radius. EANM's equation performs reasonably, but - as is clearly seen from the difference plots - it does not reproduce the linear behavior for large spheres, where differences up to 0.06 in $RC$ are shown for this equation. The linear approximation seems reasonable for $RC$=0.7-1, where $RC_{\max}$ is close to one. Equation (12) fits the $RC$-curve until $\sigma/r=1$ very closely with a difference less than 0.01. The fitted shape parameter $a$ was 2.067 ± 0.008 (uncertainty is always stated as ± one standard deviation (SD)) for spheres. All curves in Figure 2 were fitted between $\sigma/r=0$ and 2. For $\sigma/r > 1.5$ $RC$ can be approximated by the value for $RC_{\max}$.

For comparison with other results with $RC$ as a function of volume[14] and diameter[15], Figure 3 shows the $RC$ for spheres as a function of the generalized radius. The transitional area is smaller and therefore it becomes more difficult to judge the differences between the curves. This shows why it is better to plot $RC$ as a function of the reciprocal radius. Ramonaheng's equation (3) does not fit the *RC*-curve very well. The curves in Figure 3 were fitted between *R/σ*=0 and 10. From the difference plot in figure 3 (right) it is clear that only equation (10) and (12) describe the behavior for large spheres correctly, and so do the theoretical equations.

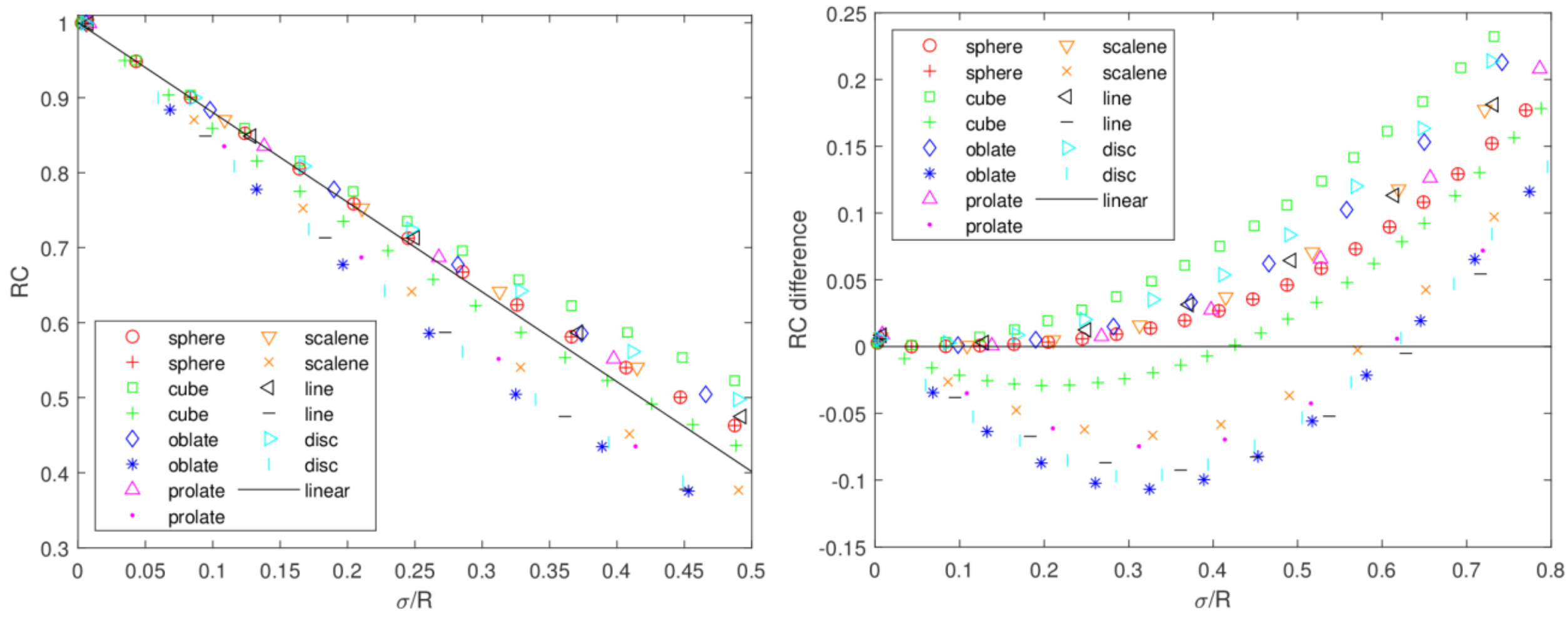

***Figure 1**: The recovery coefficient RC (left) and difference with the theoretical RC (right) for the simulated data as a function of the reciprocal radius R for the shapes in Table 1. R refers to both the generalized radius ρ (marker symbols enclosing a space such as diamonds and triangles) and volume equivalent spherical radius r (marker symbols not enclosing a space such as crosses and asterisks). The solid line corresponds to equation (10), which is linear in the reciprocal generalized radius. Spheres are effectively shown by a cross inside a circle, since for spheres ρ is equal to r. Note that the horizontal axis of the RC-difference plot goes up to 0.8, while this is 0.5 in the RC plot.*

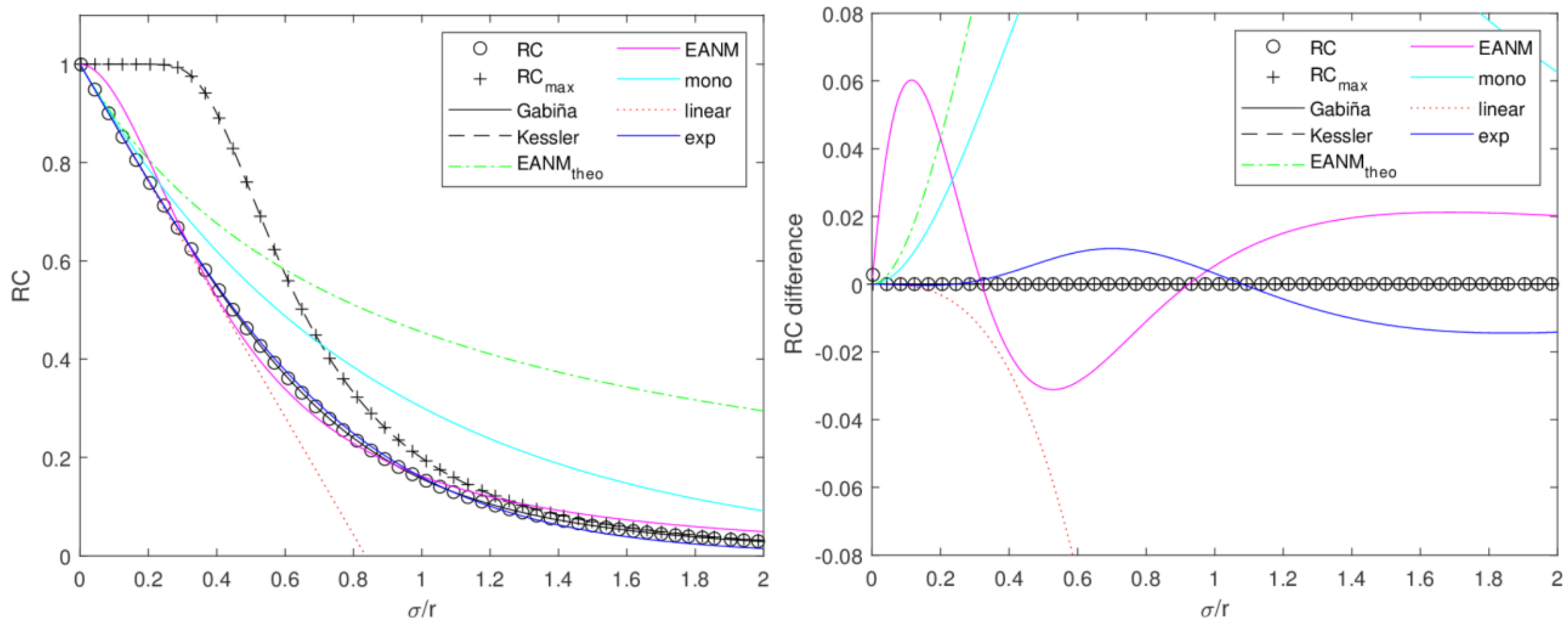

***Figure 2**: The recovery coefficient RC (left) and difference with the theoretical RC (right) as a function of the reciprocal radius r for spheres. The simulated data is labeled as black circles for $RC_{mean}$. $RC_{max}$ is the RC for the center of the sphere*[4] *and the simulated data is labeled as black crosses. The black curves are Gabiña's equation for $RC_{mean}$, Kessler's equation for $RC_{max}$, EANM's equation (4) with β=⅓ and slope according to equation (10) denoted as* $EANM_{theo}$, *EANM's two-parameter fit with equation (4), the mono-exponential curve in equation (5) with slope k=c according to equation(10), the linear approximation of equation (10) and the fit of equation (12) with the exponential labelled as "exp".*

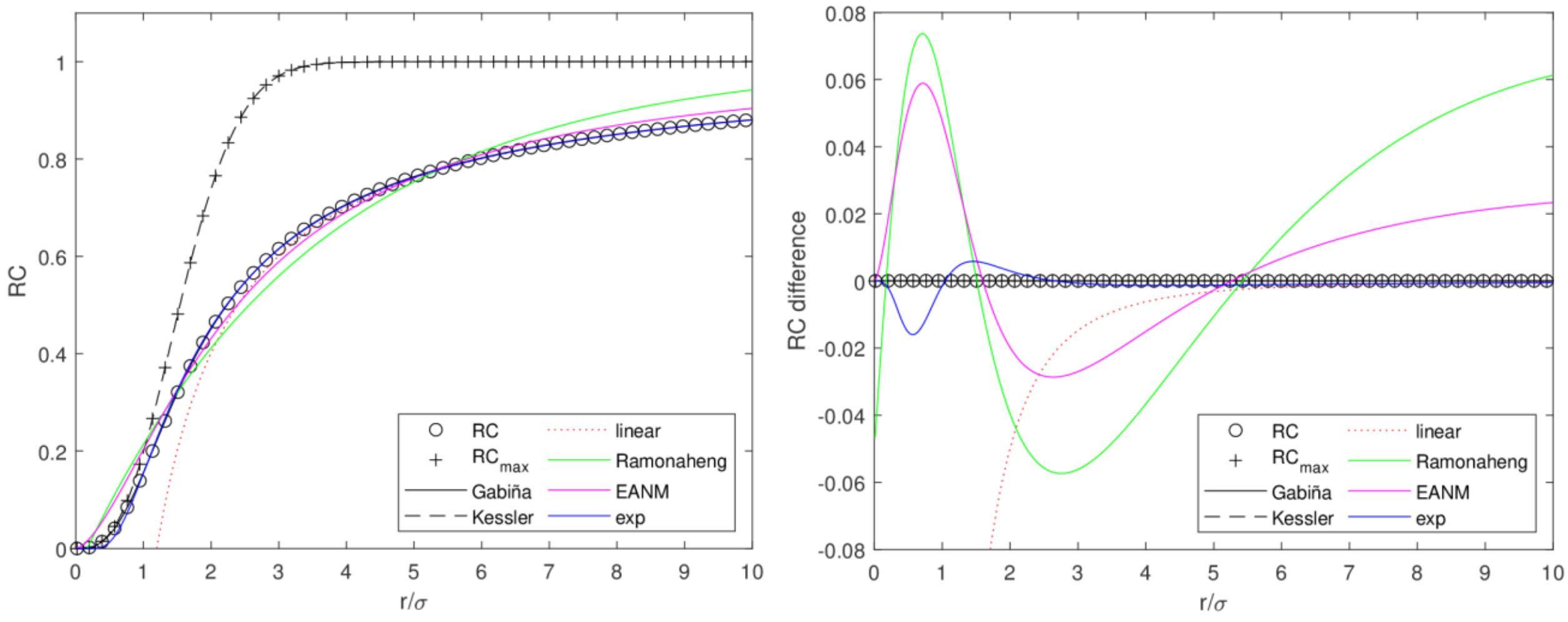

***Figure 3**: The recovery coefficient RC (left) and difference with the theoretical RC (right) as a function of the radius r for spheres. The simulated data is labeled as black circles. $RC_{max}$ is the RC of the center of the sphere*[4] *and the simulated data is labeled as black crosses. The displayed curves are Gabiña's equation for $RC_{mean}$, Kessler's equation for $RC_{max}$, the linear approximation of equation (10), EANM's two-parameter fit with equation (4), Ramonaheng's equation (3) with $b_1$=1, and the fit of equation (12) with the exponential labelled as "exp".*

In Figure 4 the recovery coefficients for spheres, cubes and discs as defined in Table 1 are shown. For the $RC_{max}$-values, there is perfect agreement between the simulated data and the theoretical curves given by Kessler et al[4] for spheres and discs (cylinders) and equation (I.3) for cubes. For *RC*-values, the linear equation (10) (dotted) and the curves fitted with equation (12) are shown. The theoretical curves for spheres and cubes are displayed in the right graph as solid curves. They are in good agreement with the numerical simulations. The fit-parameter $a$ for cubes and the discs as defined in Table 1 is $1.486 \pm 0.004$ and $1.48 \pm 0.02$, respectively. Due to overlapping the fitted $RC_{\text{mean}}$-curve for cubes is not clearly visible. Note that the linear behavior of the different shapes is identical for large objects, but not for smaller objects. The latter behavior is described by the different values of fit-parameter $a$. The difference plot in Figure 4 (right) shows the difference between the fits with equation (12) and the simulated data. For spheres and cubes, the theoretical difference is shown as a solid line with absolute difference $\lesssim 0.01$ up to $\sigma/\rho = 1.3$ ($RC \approx 0.1$). There are some small (absolute) differences ($\lesssim 0.005$) for the $RC_{\text{max}}$ for discs. The absolute differences for the $RC_{\text{mean}}$ for discs were $\lesssim 0.02$ and significantly larger than for cubes and spheres ($\lesssim 0.01$).

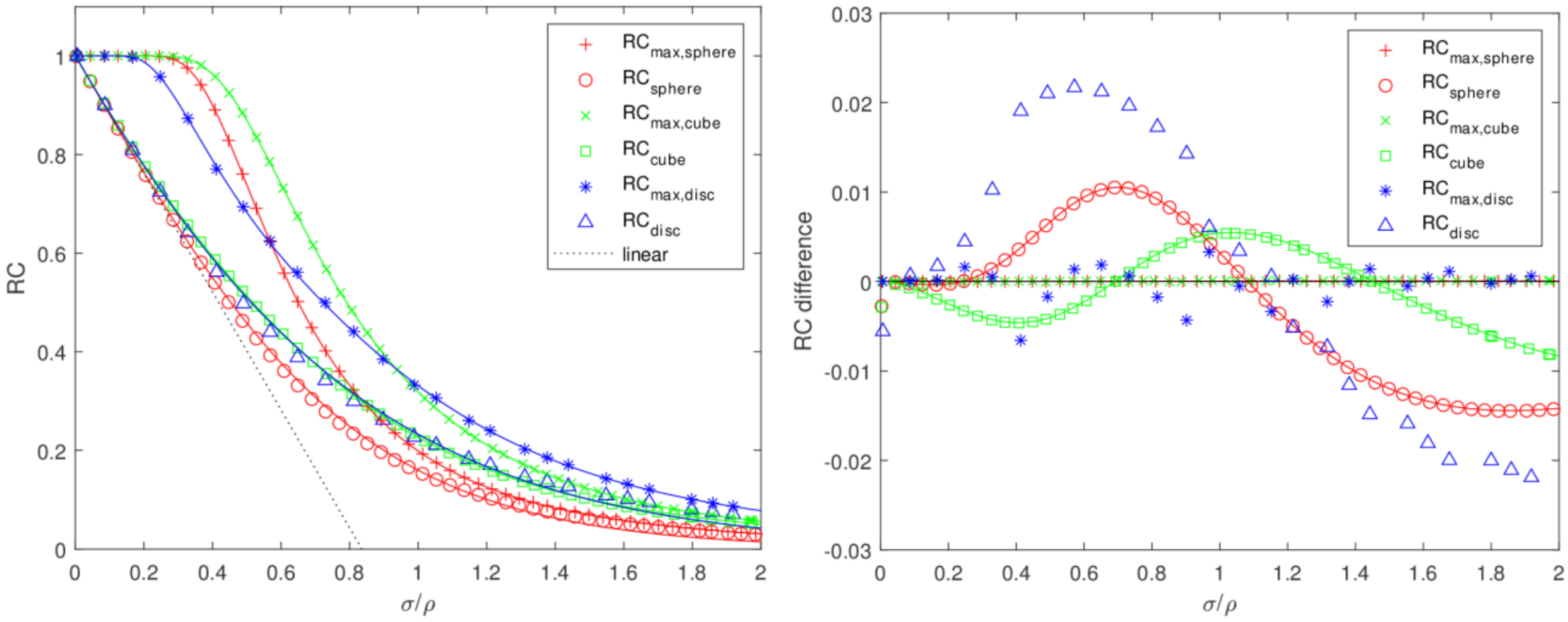

***Figure 4****: The recovery coefficient (left) as function of the reciprocal generalized radius ρ for spheres, cubes and discs, see Table 1. $RC_{max}$-curves are Kessler's equations (I.1) and (I.2) for spheres and discs (cylinders) and the $RC_{max}$-curve for cubes is given by equation (I.3). RC-curves are fitted with equation (12) and the equation (10) labeled as "linear" is shown (dotted). The differences between the fit with equation (12) and simulated RC-values are displayed in the graph to the right. For $RC_{max}$-values the difference between the theoretical values and simulated values are shown. Solid curves in*

*the graph on the right are theoretical RC- curves for cubes according to equation (I.4) and equation (I.5) for spheres, see appendix I.*

*Phantom kidney data*

Grings et al[13] published phantom data for kidneys and demonstrated the linear behavior of *RC* as function of the surface area to volume ratio. For the other non-zero background the published recovery coefficients should have been corrected for the background using equation (2) for calculation of the *CRC*. Fitting measured *RC*-values in % versus *S:V* in $cm^{-1}$ without a fixed value of 100% for zero *S:V* resulted in a coefficient of determination of $R^2$=81% and 96.9-14.8·*S:V* for the fitted linear equation. This is a significantly different result compared to the findings presented in Grings et al[13], where $R^2$=96% and 99.8-16.1·*S:V* was reported. If a 100% *RC* at zero *S:V* is fixed, the values are $R^2$=81% and 100-16.0·*S:V*. From the slope (here 0.160 cm), the system resolution can directly be calculated by multiplying with $4\sqrt{\pi \ln 2} \approx 5.90$, see equation (11), yielding FWHM=9.4mm, which is a realistic value compared to the mentioned 7-15 mm[13]. Graphs of *RC* versus *S:V* can be found in Grings et al[13].

The average sphericity (mean ± SD) for cortex-only and whole-parenchyma phantoms was calculated as 0.32 ± 0.03 and 0.61 ± 0.02, respectively. These values are significantly lower than the theoretical sphericity values in Table 1 indicating a relatively large surface area to volume ratio. Fitted and theoretical *RC*-values were plotted against the measured *RC*-values in Figure 5, which clearly demonstrates that the slope of the linear regression between the measured *RC*-values and the fitted *RC*-values is very close to one. The coefficient of determination $R^2$ for the fit with a fixed average sphericity was 88%, while it was 79% for the fit based on the measured *S:V*-ratio by Grings et al.

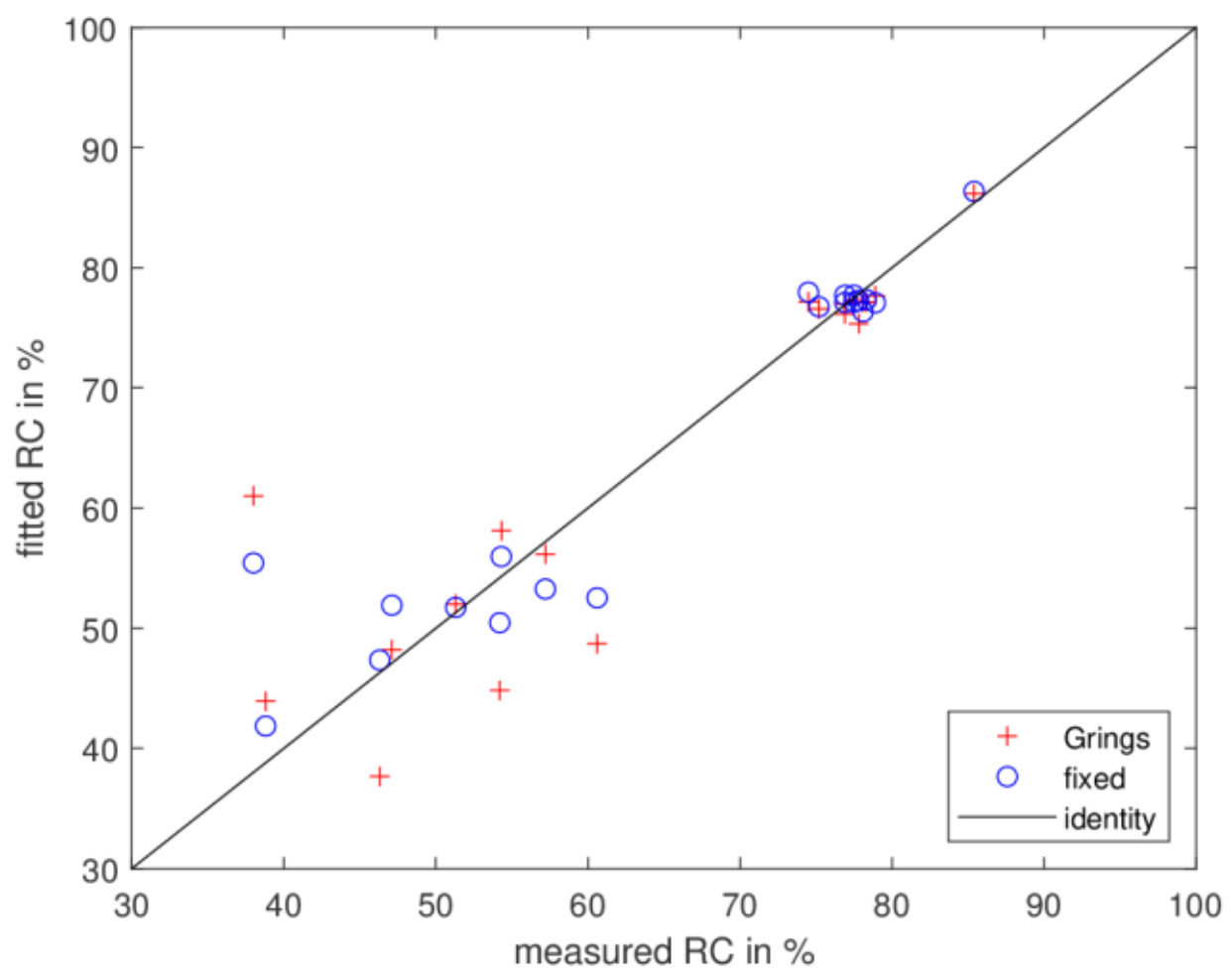

***Figure 5****: Calculated RC-values based on the fitted equation (labeled "Grings") and the fit-based equation with fixed sphericity (labeled "fixed") for the two types of kidney phantoms. Measured RC-values are displayed on the horizontal axis, and the fit-values by Grings and the fit-values based on fixed sphericity values are displayed on the vertical axis. The coefficient of determination $R^2$ for the fit by Grings was 0.79, while it was 0.88 for the fit based on fixed sphericity values. Slopes were 0.994 and 1.001, respectively. The line of identity is displayed as a black solid line.*

*EARL PET-data*

Figure 6 shows a comparison between EARL PET-data to equation (10). The four largest spheres fit very well, when a FWHM of 6 mm for EARL1 and 3 mm for EARL2 is used. The predicted inversely proportional behavior is observed, although the two smallest volumes have a *RC*, which is lower than expected by the linear fit. This is contrary to the behavior of the *RC* for small volumes in the numerical simulations.

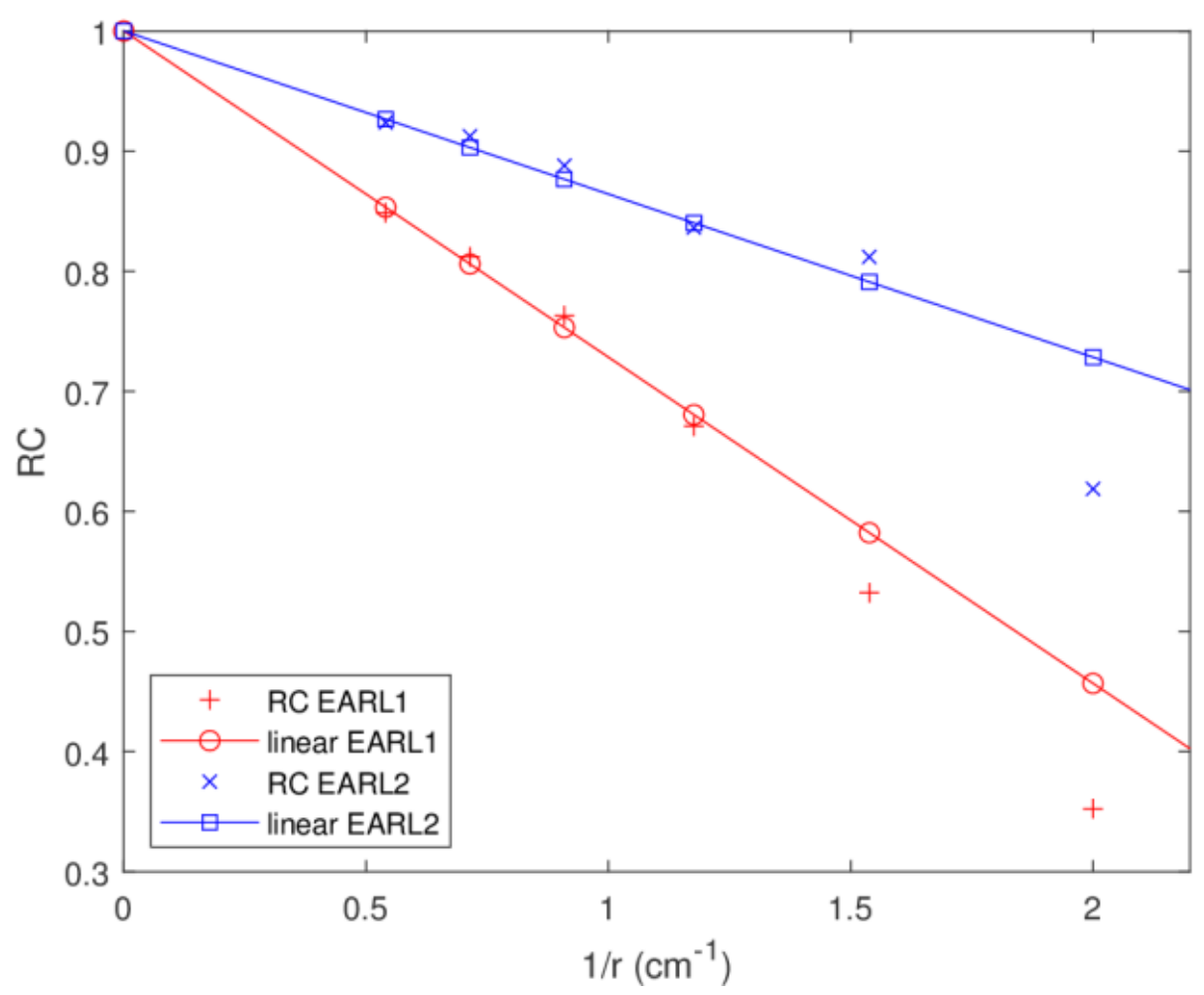

***Figure 6***: *PET EARL-data. The RC as a function of the reciprocal sphere radius for EARL1 and EARL2 are compared to the linear equation (10), where a (Gaussian) FWHM-value of 6 mm is used for EARL1 (labeled linear EARL1) and a FWHM of 3 mm for EARL1 (labeled linear EARL2).*

## Discussion

A practical rule of thumb for large objects is that the signal loss (1-*RC*) approximately (<2% deviation) equals the FWHM divided by the generalized diameter $2\rho$. In clinical practice, the generalized radius or the surface area of objects is unknown - or at least difficult to determine - while calculating the volume is more straightforward by counting the number of voxels inside the object. A practical workaround is to write the generalized radius as a fraction of the volume equivalent spherical radius $r$, which equals the sphericity $\psi$. There are some challenges related to determining the surface area of an object. It is difficult to determine it based on voxelized images, since it is not straightforward to distinguish between the total surface area of the voxels and surface area of the object. In the case of poor spatial resolution, small spatial variations (or an intricately folded surface[18]) on the object's surface are averaged out, excluding their contribution to the total surface area, which is required for the calculation of the mean recovery coefficient.

The numerical simulations in Figure 1 revealed that the linear equation (10) offers a good fit down to $RC \approx 0.7$ for the investigated shapes (absolute difference $\lesssim 0.01$), and they confirm the necessity of including the sphericity as a shape factor. For smaller objects, the *RC*-curve starts to bend upwards.

The only equation that reproduces both the linearity for large objects and the bending of the curve (until $RC \approx 0.1$) is equation (12). Equation (3) from the EANM guidelines results in poor curves, when linearity for large objects is demanded by setting $\beta$=⅓ and the theoretical slope from equation (10) is used. Fitting these two parameters yields physically less meaningful parameters and a better fit, but it does not reproduce the linearity for large objects and results in *RC*-values that are too high for $RC < 0.25$. Fitted values for $\beta$ are typically larger as for example for spheres in Tran-Gia et [12], where 0.6 ± 0.2 (mean ± SD) was found. The numerical simulations yield 0.64 ± 0.01 for spheres fitted between 0 and 2 for $\sigma/r$.

In the MIRD pamphlet No. 23[24], it is suggested that the sphere model can be used (with less than 10% inaccuracy) for ellipsoids with major-to-minor axis ratios of less than two, by using the *RC* for volume equivalent spheres[25], and thus the volume equivalent radius. A better approach would be to use the generalized radius in order to take the sphericity into account. The sphericity for an oblate ellipsoid (lowest sphericity of the three different types) with ratio two is 0.91. This confirms the degree of inaccuracy and the necessity of taking the sphericity into account.

Theoretical curves for $RC_{\max}$ fit well for spheres, cubes and discs (cylinders), and demonstrate different behavior depending on the shape. Spheres feature the steepest transition, while discs feature the flattest transition. Curves for mean *RC*-values also reveal different behavior for small objects, resulting in a different fit-parameter $a$ in equation (12) depending on the shape, but identical linear behavior as a function of the reciprocal generalized radius is conserved for large objects.

Remarkably, the linear fit with fixed sphericity for the two types of kidney phantoms result in a smaller spread than the linear fit with measured sphericity. This can be seen in Figure 5 and might be due to uncertainty or inaccuracies in the surface area values. Thus, in the case of imaging organs such as kidneys, the equation allows for a solid approximation based on a fixed sphericity and the use of a size (e.g. volume) based *RC*. From the original kidney phantom data[13], it is clear that the background is not taken into account. The target-to-background ratio should be utilized to change *RC* into *CRC* according to equation (2). The higher value for the coefficient of determination $R^2$ for the *RC*-curve in Grings et al[13] is possibly caused by at least one missing data point in their Figure 5 as well as possibly their fit. The data point "Pat3R" has 2.4 $cm^{-1}$ for the surface area to volume ratio and 38.0% for the recovery coefficient as can be found in their Table 1 and 2, respectively. This data point is not shown in Figure 5. The data point "Pat1R" with $RC$=60.6% and $S$:$V$=3.2 $cm^{-1}$ cannot be seen in their Figure 5 either and a displayed data point with an $RC$ just above 60% and a $S$:$V$ of approximately 2.5 $cm^{-1}$ for the cortex only phantom with the lowest $S$:$V$ cannot be found in their Table 1 and 2.

For the EARL-data a good fit was achieved with a FWHM of 6 mm for EARL1 and 3 mm for EARL2 for all but the two smallest spheres, which had an unexpected lower *RC*-value. A possible explanation is the limited signal-to-noise ratio, which causes a loss of signal and poorer spatial resolution.

All the results discussed in this article are obtained for a uniform activity distribution (inside the objects), which is present in phantoms, but typically not in human subjects. The influence of the activity distribution itself needs to be investigated. For large objects, it is expected that with smoother activity concentration transitions between object and background, the mean $RC$ will be higher (due to less net spill-out) and the maximum $RC$ will be lower (due to immediate smoothing for small values of the FWHM).

For a square and triangularly shaped PSF, it can be derived that the numerical value of 1.0165 in equation (11) changes into 0.75 and 1, respectively, suggesting that the exact shape of the PSF is of lesser importance. This is also understandable, since the PSF is integrated.

Theoretically, equation (10) for arbitrary shapes is identical to equation (9) in Gabiña et al[18], which was derived for large spheres and experimentally found to describe ellipsoids as well. Figure 6 in Gabinã et al contains experimental data for $^{99m}Tc$ and $^{177}Lu$ and is similar to Figure 1, 2 and 4. It also demonstrates the linear behavior (for large objects) of the mean $RC$ as a function of reciprocal generalized radius.

*Conclusions*

When fitting *RC*-curves, equation (12) should be applied as a fit. For *RC*-values over 70% the linear approximation of equation (10) can be used and - if an error of three percentage points is acceptable - even down to 50%. If the spatial resolution is known, only one empirical fit-parameter is required. The other fit-parameter is directly related to the spatial resolution. For non-spherical shapes, the sphericity is incorporated in the equation - which is often far less subject dependent than size - and in those cases a fixed intersubject sphericity can be utilized. This article has demonstrated that the novel equation and the linear approximation work well for both simulations and a variety of shapes, as well as in practice for phantoms containing spheres in PET and kidney shapes in SPECT. It can be applied in dosimetry, partial volume correction, as well as possibly in improved resolution modeling and for the calculation of resolution independent standard uptake values.

## Acknowledgements

The author would like to thank Sune H. Keller and Claus Svarer for the fruitful discussions, and for carefully reading the manuscript.

***Table 1****: Generalized radii and sphericity values for a sphere, cube, disc, line and ellipsoids according to equation (9), where r is the volume equivalent radius, S the surface area and V the volume. A shape parameter q is used as the fraction of the smallest and largest dimension.*

<table>
<tr><td>parameter<br>shape<br>$0<q<1$</td><td>generalized radius<br>$\rho = \frac{3V}{S} = \psi r$</td><td>sphericity<br>$\psi = \frac{\sqrt[3]{36\pi V^2}}{S} = \frac{\rho}{r}$</td><td>sphericity<br>$\psi, q = \frac{1}{4}$ *</td></tr>
<tr><td>sphere<br>radius $R$</td><td>$R$</td><td>1</td><td>1</td></tr>
<tr><td>cube<br>edge length $d$</td><td>$d/2$</td><td>$(\pi/6)^{1/3}$</td><td>0.8060</td></tr>
<tr><td>disc<br>diameter $D$<br>height $h=qD$</td><td>$\frac{3qD}{2+4q}$</td><td>$\left(\frac{9}{4}\right)^{1/3} \frac{q^{2/3}}{\frac{1}{2}+q}$</td><td>0.6934</td></tr>
<tr><td>line (cylinder)<br>length $L$<br>diameter $D=qL$</td><td>$\frac{3qL}{4(1+\frac{q}{2})}$</td><td>$\left(\frac{9}{4}\right)^{1/3} \frac{q^{1/3}}{1+\frac{q}{2}}$</td><td>0.7338</td></tr>
<tr><td>oblate ellipsoid<br>$q$:1:1</td><td colspan="2" rowspan="3">approximated[21] by (error in $S$ <1.2%)<br>$S \approx 4\pi \left(\frac{(ab)^p+(ac)^p+(bc)^p}{3}\right)^{\frac{1}{p}}$ with $p = \frac{8}{5}$<br>$V = \frac{4\pi}{3} abc$,<br>where $a$, $b$ and $c$ are the principal semi-axes.</td><td>0.6972</td></tr>
<tr><td>prolate ellipsoid<br>$q$:$q$:1</td><td>0.7852</td></tr>
<tr><td>scalene ellipsoid<br>$q$:2$q$:1</td><td>0.7915</td></tr>
</table>

* if applicable

## Appendix I

In Kessler et al[4], analytical equations for $RC_{\max}$ are given for a sphere with radius $R$ and a cylinder with radius $R$ and height $H$:

$$RC_{\max}^{\text{sphere}} = \text{erf}\left(\frac{R}{\sigma\sqrt{2}}\right) - \sqrt{\frac{2}{\pi}}\frac{R}{\sigma}\,\mathrm{e}^{-\frac{R^2}{2\sigma^2}}, \tag{I.1}$$

$$RC_{\max}^{\text{cylinder}} = \left(1 - \mathrm{e}^{-\frac{R^2}{2\sigma^2}}\right)\text{erf}\left(\frac{H}{\sigma\cdot 2\sqrt{2}}\right). \tag{I.2}$$

Note the uncommon definition of the error function[1] in Kessler et al. An equation for a cube with edge length $d$ can be derived by integration of the point spread function. The center of the cube is positioned in the origin of a 3D space with Cartesian coordinates $x$, $y$ and $z$. The equation for $RC_{\max}$ for a cube is given by

$$RC_{\max}^{\text{cube}} = \iiint_{\text{cube}} \left(\frac{1}{\sigma\sqrt{2\pi}}\right)^3 \mathrm{e}^{-\frac{x^2+y^2+z^2}{2\sigma^2}} dxdydz = \left(\frac{1}{\sigma\sqrt{2\pi}}\int_{-\frac{d}{2}}^{\frac{d}{2}} \mathrm{e}^{-\frac{x^2}{2\sigma^2}} dx\right)^3 = \left(\text{erf}\left(\frac{d}{\sigma\cdot 2\sqrt{2}}\right)\right)^3, \tag{I.3}$$

where a 3D Gaussian PSF is used. According to equation (8) and (14) in Di Martino et al[19], the mean $RC$ for a cube is given by

$$RC_{\text{mean}}^{\text{cube}} = \left(\text{erf}\left(\frac{d}{\sigma\cdot\sqrt{2}}\right) - \sqrt{\frac{2}{\pi}}\frac{\sigma}{d}\left(1 - \mathrm{e}^{-\frac{d^2}{2\sigma^2}}\right)\right)^3, \tag{I.4}$$

where the numerator inside the erf-function has been corrected from $d/2$ to $d$. Equation (6) and (14) in di Martino et al can be used to evaluate the $RC$ in the center, which equals $RC_{\max}$, and is in agreement with equation (I.3). For large cubes ($d \rightarrow \infty$), equation (I.4) is equal to equation (10).

Gabiña et al[18] derived an equation for the mean $RC$ for spheres. It is given by

$$RC_{\text{mean}}^{\text{sphere}} = \text{erf}\left(\frac{R\sqrt{2}}{\sigma}\right) - \frac{1}{\sqrt{2\pi}}\frac{\sigma}{R}\left(3 - \mathrm{e}^{-\frac{2R^2}{\sigma^2}}\right) + \frac{1}{\sqrt{2\pi}}\left(\frac{\sigma}{R}\right)^3\left(1 - \mathrm{e}^{-\frac{2R^2}{\sigma^2}}\right). \tag{I.5}$$

For large spheres ($R \rightarrow \infty$) this equation is equal to equation (10) as well.

---

[1] Here the error function is defined as $\text{erf}(x) = \frac{2}{\sqrt{\pi}}\int_0^x \mathrm{e}^{-t^2} dt$.

## Appendix II

Equation (8) for the smoothing thickness $\delta$ can be evaluated by substitution with $z = \frac{x'-x}{\sigma\sqrt{2}}$ :

$$\delta = \frac{1}{\sigma\sqrt{2\pi}} \int_{-\infty}^{0} \int_{0}^{\infty} e^{-\frac{(x-x')^2}{2\sigma^2}} dx' \, dx = \frac{1}{\sqrt{\pi}} \int_{-\infty}^{0} \int_{-\frac{x}{\sigma\sqrt{2}}}^{\infty} e^{-z^2} dz \, dx \qquad \text{(II.1)}$$

The inner integral can be written as the sum of two error functions:

$$= \frac{1}{\sqrt{\pi}} \int_{-\infty}^{0} \left\{ \int_{-\frac{x}{\sigma\sqrt{2}}}^{0} e^{-z^2} dz + \int_{0}^{\infty} e^{-z^2} dz \right\} dx = \frac{1}{2} \int_{-\infty}^{0} \left\{ \operatorname{erf}\left(\frac{x}{\sigma\sqrt{2}}\right) + 1 \right\} dx \qquad \text{(II.2)}$$

Substitution with $y = \frac{x}{\sigma\sqrt{2}}$ yields:

$$= \frac{\sigma}{\sqrt{2}} \int_{-\infty}^{0} \{\operatorname{erf}(y) + 1\} dy = \frac{\sigma}{\sqrt{2}} \left[ y \cdot \operatorname{erf}(y) + \frac{e^{-y^2}}{\sqrt{\pi}} + y \right]_{-\infty}^{0} = \frac{\sigma}{\sqrt{2\pi}} \qquad \text{(II.3)}$$

## Appendix III

The maximum and minimum of the FWHM for the numerical simulations are derived. In order to obtain an accurate convolution, it was required that the distance between the object and matrix border was three $\sigma$, which is denoted as $c_\sigma \sigma$, so that the tail of the Gaussian is close to zero at the matrix border. The maximum *σ/R* or FWHM*/L* for an object of size *L* and Field of View size *FoV* is found by assuming that *L* equals the pixel size *FoV/n* with *n* the number of pixels, i.e. the matrix size of the image. This yields:

$$FoV = L + 2c_\sigma \sigma = L + Lc_\sigma \cdot \frac{\text{FHWM}}{\sqrt{2\ln 2}L} \quad \rightarrow \quad \frac{\text{FWHM}_{\text{max}}}{L} = \sqrt{2\ln 2} \cdot \frac{n-1}{c_\sigma} \qquad \text{(III.1)}$$

The minimum FWHM (system spatial resolution) is determined in a similar way by assuming that the pixel size is equal to $c_\sigma \sigma$ yielding:

$$FoV = L + 2\frac{FoV}{n} \quad \rightarrow \quad \frac{FoV}{n} = \frac{L}{n-2} = c_\sigma \sigma \quad \rightarrow \quad \frac{\text{FWHM}_{\text{min}}}{L} = \frac{\sqrt{8\ln 2}}{(n-2)c_\sigma} \qquad \text{(III.2)}$$